\documentclass[12pt]{article}
\def\double{\baselineskip 24pt \lineskip 10pt}

\begin{document}
\begin{center}
\vspace{.8cm}
\Large
{\bf GRAVITATION IS TORSION}\\

\vspace{.5cm}
\vspace{.5cm}

\large{Engelbert Schucking} \\
\vspace{.15cm}
\normalsize
Physics Department \\
New York University \\
4 Washington Place \\
New York, NY 10003 \\
els1@nyu.edu \\

\vspace{.8cm}

\end{center}

\double

\begin{abstract}\double
The mantra about gravitation as curvature is a misnomer. The     
curvature tensor for a standard of rest does not describe acceleration in a gravitational field but the \underline{gradient} of the acceleration (e.g. geodesic deviation).
The gravitational field itself (Einstein 1907) is essentially an accelerated reference system. It is characterized by a field of ortho-normal four-legs in a Riemann space with Lorentz metric.
By viewing vectors at different events having identical leg-components as parallel (teleparallelism) the geometry in a gravitational field defines torsion. This formulation of Einstein's 1907 principle of equivalence uses the same Riemannian metric and the same 1916 field equations for his theory of gravitation and fulfills his vision of General Relativity.
\end{abstract}

\eject

\section{Why Gravity is \underline{not}
\lq\lq Manifestation of Space-Time Curvature\rq\rq.$^{1}$}
 
By gravity one understands the force that draws apples toward the Earth.
Its field strength near the surface of our planet is characterized by
a radial acceleration $ g = 9.8 m/sec^{2} $ that it imparts to all matter.

     The curvature of the pseudo-Riemannian space-time  manifold is described by its Riemann tensor that generalizes the intrinsic curvature of 2-dimensional surfaces defined by Carl Friedrich Gauss.

     To investigate the relation between gravitation and curvature in General Relativity I shall discuss the falling of a mass point (e.g. an idealized apple) near the surface of an idealized spherical non-rotating Earth.
If the mass point falls radially its motion takes place in the space-time \lq\lq plane\rq\rq\ spanned by a radial coordinate $r$ and a time coordinate $t$.

     Karl Schwarzschild and Johannes Droste discovered the geometry in this plane by finding the exact solution of Einstein's vacuum field equations for a spherically symmetric gravitational field. They wrote for the square of the line element ds
\begin{equation}
ds^2 \, = \, \biggr( 1 - \frac{2m}{r} \biggr) \, (d(ict))^2  \, + \,
\biggr( 1 - \frac{2m}{r} \biggr)^{-1} \, (dr)^2 \, .
\end{equation}
In this formula $ds$ describes the infinitesimal distance between two points
with coordinates $(t, r)$ and $(t + dt, r + dr)$. The radial distance $r$ is fixed by
stating that the area $A$ of a concentric sphere with radius $r$ is given by
Archimedes' formula $ A = 4 \pi r^2 $. The constant $m$ stands for
\begin{equation}
m \, = \, G \, \frac{M}{c^2}
\end{equation}
where $G$ is Newton's gravitational constant, $M$ the mass of the Earth and
$c$ the speed of light in vacuo. Finally, $i$ is short for $ \sqrt{-1} $.

I use the structural equations$^2$ of \'Elie Cartan for calculation of the Gaussian curvature K in the $t$-$r$-plane. In a $n$-dimensional Riemann space Cartan writes the square of the line element $ds$ in terms of n differential one-forms $\omega_j$ with the index $j$ running from $0$ to $n-1$
\begin{equation}
ds^2 \, = \, (\omega_0)^2 \, + . . . + \, (\omega_{n-1})^2 \, .
\end{equation}
The first structural equation says
\begin{equation}
\Theta_j \, = \, d\omega_j \, + \, \omega_{jk} \wedge \omega_k \, .
\end{equation}
Here $\Theta_j$ are the torsion two-forms while $\omega_{jk}$ are the connection one-forms.
Einstein summation over the repeated index $k$ is implied and the symbol $\wedge$
denotes the skew-symmetric multiplication of differential forms.
For a metric connection that preserves length under parallel transport we have
\begin{equation}
\omega_{jk} \, = \, - \, \omega_{kj} \, .
\end{equation}

In our case $ n = 2 $ we have from (1) and (3) the two differential forms
\begin{equation}
\omega_0 \, = \, \biggr( 1 - \frac{2m}{r} \biggr)^{1/2} \, d(ict) , \ \ \ \
\omega_1 \, = \, \biggr( 1 - \frac{2m}{r} \biggr)^{-1/2} \, dr \, .
\end{equation}
If we assume that the torsion forms vanish: $ \Theta_j = 0 $, the connection form 
$ \omega_{01} = - \omega_{10} $ defines the Levi-Civita connection subject to the two equations from (4) 
\begin{equation}
0 \, = \, d\omega_0  +  \omega_{01} \wedge \omega_1 \, , \ \ \ \
0 \, = \, d\omega_{1} - \omega_{01} \wedge \omega_0 \, .
\end{equation}
According to (6) the differential two-form $ d\omega_1 $ vanishes.
The second equation (7) tells us that the connection form $ \omega_{01} $
must be proportional to the one-form $ \omega_0 $,
say, $ \omega_{01} = \lambda \omega_0 $,
with some scalar function $\lambda$.
Inserting this relation into the first equation (7) gives, with (6),
\begin{equation}
0 \, = \, \frac{m}{r^2} \, \biggr( 1 - \frac{2m}{r} \biggr)^{-1/2} dr \wedge d(ict) \, + \,
     \lambda \, d(ict) \wedge dr \, .
\end{equation}
We read off that 
\begin{equation}
\lambda \, = \,  \frac{m}{r^2} \, \biggr( 1 - \frac{2m}{r} \biggr)^{-1/2} \, .
\end{equation}
The connection form $ \omega_{01} $ becomes
\begin{equation}
\omega_{01} \, = \,  \lambda \, \omega_0
\, = \, \frac{m}{r^2} \, \biggr( 1 - \frac{2m}{r} \biggr)^{-1/2} \omega_0
\, = \, \frac{m}{r^2} \, d(ict) \, .
\end{equation}
The curvature follows from Cartan's second structural equation giving
for $ n = 2 $ the curvature two-form $ \Omega_{01} $
\begin{equation}
\Omega_{01} \, = \, R_{0101} \ \omega_0 \wedge \omega_1
\, = \, d\omega_{01} \, = \, \frac{2m}{r^3} \ \omega_0 \wedge \omega_1 
\, = \, K \, \omega_0 \wedge \omega_1 \, .
\end{equation}
By (11) the curvature form defines the component $ R_{0101} $ of the Riemann tensor
and the Gaussian curvature $K$
\begin{equation}
R_{0101} \, = \, K \, = \, \frac{2m}{r^3} \, .
\end{equation}

For a discussion of the connection form $ \omega_{01} $,
I introduce the reference frame $ {\bf e}_k $.
It consists of $n$ ortho-normal vector fields $ {\bf e}_k $
tangent to the manifold that are dual to Cartan's differential one-forms $ \omega_j $
\begin{equation}
\omega_j ({\bf e}_k) \, = \, \delta_{jk}
\end{equation}
where Kronecker's $ \delta_{jk} $ is the unit matrix.
Katsumi Nomizu's process of covariant differentiation $ \nabla_{\bf v} {\bf w} $ 
of a vector field $ {\bf w} $ with respect to a tangent vector $ {\bf v} $ 
gives for $ {\bf w} = {\bf e}_k $
\begin{equation}
\nabla_{\bf v} {\bf e}_k \, = \, {\bf e}_j \, \omega_{jk}({\bf v})
\end{equation}
defining the connection one-forms $ \omega_{jk} $.
The covariant derivative of the time-like vector field $ {\bf e}_0 $
with respect to $ {\bf v} = {\bf e}_0 $ defines
the negative of the geodesic acceleration  $ - a \, {\bf e}_1 $
\begin{equation}
\nabla_{{\bf e}_0} {\bf e}_0 \, = \,
 - \, {\bf e}_1 \, \omega_{01}({\bf e}_0) \, = \,
 - \, \lambda \, {\bf e}_1 \, = \, - a \, {\bf e}_1
\end{equation}
where $a$ is the magnitude of the geodesic acceleration. This equation shows that 
\begin{equation}
a \, = \,
\lambda \, = \,
\biggr( \frac{m}{r^2} \biggr) \biggr( 1 - \frac{2m}{r} \biggr)^{-1/2} \, .
\end{equation}

Calculating the radial gradient of the geodesic acceleration $ da/ds $, we obtain
\begin{equation}
\frac{da}{ds} = \biggr(1 - \frac{2m}{r}\biggr)^{1/2} \, \frac{da}{dr}
\, = \, - \, \biggr( \frac{2m}{r^3}\biggr)
    \biggr( 1 - \frac{3m}{2r} \biggr)
    \biggr(1 - \frac{2m}{r}\biggr)^{-1} \, .
\end{equation}

     When the radius $r$ is large compared to the Schwarzschild radius $ 2m $
the geodesic acceleration a becomes Newton's $ g/c^2 $ and $ da/ds $ its gradient. 
We identify thus the geodesic acceleration with the gravitational field strength and the space-time curvature $ 2m/r^3 $ with its gradient. While the geodesic acceleration has dimension of inverse length, the space-time curvature has dimension of inverse length squared and is thus not suitable to manifest gravity. 
Even the square root of the curvature, a sort of \lq\lq radius of curvature\rq\rq, bears no order of magnitude relation to the geodesic acceleration as one can easily infer by putting numbers into the equations for the surface of the Earth. A drastic proof of this fact is the simple observation that by letting $m$ and $r$ going to infinity in (16), but keeping the geodesic acceleration a fixed, one can have arbitrary acceleration for vanishing curvature.
This shows that there are gravitational fields of arbitrary strength in Minkowski space-time.

The statement that gravity is a manifestation of space-time curvature is misleading. And misleading too are the illustrations in physics texts, or models in exhibitions, purporting to explain Einstein's theory of gravitation by having balls orbiting on curved surfaces as fake planets or satellites. Equally flawed are the \lq\lq proofs\rq\rq\ in current texts that the gravitational red-shift near the Earth's surface shows that space-time is curved.

\section{The Geometry of the Gravitational Field.$^{3}$}

In 1928 Einstein discovered distant parallelism.
This is the geometry of the gravitational field as confirmed by
the Pound-Rebka experiment in 1960. 
Riemann's geometry can be interpreted in terms of distant parallelism as follows:
Above, we had set the torsion forms $ \Theta_j $ equal to zero
for determining the $ \omega_{jk} $ of the Levi-Civita connection.
We can write instead
\begin{equation}
\Theta_j \, = \, - \, \omega_{jk} \wedge \omega_k \, = \, d\omega_j
\end{equation}
for Einstein's distant parallelism that has torsion but vanishing connection forms. Here vectors are now considered to be equal and parallel if they have the same components with respect to the given ortho-normal frame.
In particular, the frame vector ${\bf e}_0$ becomes now tangent to  time-like world lines that provide the standard of rest in the gravitational field. Time is now no longer measured along geodesics.

In a rectangle of the four points $(r_1, t_1)$, $(r_1, t_2)$, $(r_2, t_1)$, $(r_2, t_2)$
its opposite time-like sides show, according to (1), a ratio of length
\begin{equation}
\biggr[\frac{(1 - 2m/r_2)}{(1 - 2m/r_1)}\biggr]^{1/2}
\, \approx \,  1  +  m \, \frac{r_2 - r_1}{r_1{}^2} \, .
\end{equation}
This failure of a metric rectangle to close is precisely the result of the
Pound-Rebka experiment. It also demonstrates that the geometry of a gravitational field can be described by distant parallelism (torsion)
without appealing to curvature. Curvature terms on the left hand side of (19) are of higher order in $r_1$.

     The interpretation of Einstein's theory in terms of distant parallelism
uses the same metric, variational principle and field equations  but gives a clear definition of a gravitational field thus providing a satisfying account of General Relativity, a vision that had shriveled to not much more than \lq\lq Einstein's Field Equations\rq\rq.


\end{document}